\magnification=1200 

\headline{\ifnum\pageno=1 \nopagenumbers 
\else \hss\number \pageno \fi} 
 
\overfullrule=0pt
\footline={\hfil}
\font\boldgreek=cmmib10
\textfont9=\boldgreek
\mathchardef\myphi="091E
\def\bfphi{{\fam=9 \myphi}\fam=1}
\def\lsim{\raise0.3ex\hbox{$<$\kern-0.75em\raise-1.1ex\hbox{$\sim$}}}
\def\gsim{\raise0.3ex\hbox{$>$\kern-0.75em\raise-1.1ex\hbox{$\sim$}}}
\baselineskip=10pt 
\vbox to 1,5truecm{}
\parskip=0.2truecm 
\centerline{\bf REGGE BEHAVIOUR AND REGGE TRAJECTORY}
\medskip
 \centerline{\bf FOR LADDER GRAPHS IN SCALAR $\bfphi^{\bf 3}$ FIELD THEORY}\bigskip 

\bigskip \centerline{by}\smallskip 
\centerline{\bf R. Hong Tuan}  
\medskip
 \centerline{Laboratoire de Physique Th\'eorique et Hautes Energies
\footnote{*}{Laboratoire associ\'e au Centre National de la Recherche
Scientifique - URA D0063}}  \centerline{Universit\'e de Paris XI, b\^atiment 211, 91405
Orsay Cedex, France}  
\bigskip\bigskip
\baselineskip=20pt 
\noindent 
${\bf Abstract}$ \par 
Using the gaussian representation for propagators (which can be proved to be exact in
the infinite number of loops limit) we are able to derive the Regge behaviour for
ladder graphs of $\phi^3$ field theory in a completely new way. An analytic expression
for the Regge trajectory $\alpha (t/m^2)$ is found in terms of the mean-values of the
Feynman $\alpha$-parameters. $\alpha (t/m^2)$ is calculated in the range $- 3.6 < t/m^2
< 0.8$. The intercept $\alpha (0)$ agrees with that obtained from earlier calculations
using the Bethe-Salpeter approach for $\alpha (0) \gsim 0.3$. \par

\vbox to 4 truecm{}

\noindent LPTHE Orsay 96/46 \par 
\noindent June 1996

\vfill\supereject
The first attempts at calculating ladder graphs and their asymptotic behaviour, ($s \to
\infty$, $t$ fixed) i.e. Regge behaviour with leading Regge trajectory $\alpha (t)$,
date back to more than thirty years ago. Used were the Feynman $\alpha$-parameter
representation$^1$, the Bethe-Salpeter$^{2,4}$ equation or the multiperipheral
model$^3$ which gives a recursion equation close to the Bethe-Salpeter one.
Nakanishi$^4$ derived an exact expression for the intercept $\alpha (0)$ of the leading
Regge trajectory assuming that a massless scalar was exchanged in the central
propagators joining the sides of ladder diagrams. More recently, ladders diagrams in
massless $\phi^3$ were given an expression in terms of polylogarithms$^5$, a leading
log approximation$^6$ used to obtain the QCD Pomeron trajectory and the asymptotics of
$\phi^3$ ladders in six dimensions studied for large $t$ $^7$. \par

Here, we want to exploit the $\alpha$-parameter representation for the Feynman ladder
graphs in massive scalar $\phi^3$ theory at $d = 4$ dimensions to obtain, first, the
Regge behaviour and, second, the leading Regge trajectory $\alpha (t)$. Our approach,
however, will be very different from all those tried before. Some time ago, we gave a
proof$^{\, 8}$ that when there are an infinite number of them, the $\alpha$-parameters
can be replaced by their mean-values, once an overall scale has been extracted, and this
for superrenormalizable scalar field theories. Let us define a homogeneous polynomial
$P_G(\{\alpha \})$
$$P_G(\{\alpha \}) = \sum_{{\cal T}} \prod_{\ell \not\subset {\cal T}} \alpha_{\ell}
\eqno(1)$$
\noindent which is a sum over all spanning trees ${\cal T}$ belonging to a one-particle
irreducible graph $G$ (a spanning tree is a tree incident with all vertices of $G$).
$P_G(\{\alpha \})$ is of degree $L$ in the $\alpha_{\ell}$'s if $G$ has $L$ loops.
Similarly, let us define also
$$Q_G (P, \{\alpha \}) = P_G^{-1} (\{\alpha \}) \sum_C s_C \prod_{\ell \subset C}
\alpha_{\ell} \eqno(2)$$
\noindent where we have summed $\sum\limits_C$ over all cuts $C$ on $G$ (a cut $C$ is the
complement of a tree ${\cal T}$ on $G$ plus one propagator. This propagator cuts ${\cal
T}$ in two disjoint parts). Then, in (2) $\sum\limits_C s_C \prod\limits_{\ell \subset C}
\alpha_{\ell}$ is a homogeneous polynomial of degree $L + 1$ in the $\alpha_{\ell}$'s.
Then,
$$s_C = \left ( \sum_{v \in G_1} P_v \right )^2 = \left ( \sum_{v \in G_2} P_v \right
)^2 \eqno(3)$$
\noindent is a Lorentz-invariant quadratic in the external momenta $P_v$ associated
with the external lines of $G_1$ (or $G_2$) if $G_1$ (or $G_2$) is one part of $G$.
$P_G(\{ \alpha \})$ and $Q_G(P, \{\alpha \})$ {\it contain all the information about
the topology of $G$}. Let us define by $F_G$ the Euclidean amplitude for $G$, then for
$I, L \to \infty$, ($I$ is the number of propagators of $G$) coupling $\gamma = - 1$, we
get $$F_G = (4 \pi )^{-dL/2} h_0 \left [ P_G \{\bar{\alpha}\} \right ]^{-d/2} \left [ Q_G
(P, \{\bar{\alpha} \}) + m^2 h_0 \right ]^{-(I -d/2L)}$$
$$\Gamma (I - d/2L) \ h_0^{I-1}/(I - 1)~! \eqno(4)$$
\noindent where $h_0 = \sum\limits_{\ell} \alpha_{\ell}$ and $h_0^{I-1}/(I-1)!$ is the
phase-space volume of the $\alpha_{\ell}$'s. The factor $\Gamma (I - d/2L)$. $[Q_G(P,
\{\bar{\alpha}\}) + m^2 h_0]^{-(I-d/2L)}$ coming from the integration over the
overall scale is convergent for superrenormalizable theories (such as $\phi^3$ in $d =
4$). The next step will consist in evaluating the mean-values $\bar{\alpha}_{\ell}$.
Then, the result will be put in (4) to obtain the Regge behaviour, taking into account
the peculiar form of $P_G(\{\bar{\alpha}\})$ and $Q_G(P, \{\bar{\alpha}\})$ for ladder
graphs. \par

To calculate the mean-value $\bar{\alpha}_i$ of a Feynman parameter $\alpha_i$ attached
to a propagator $i$ one has to isolate the dependence of $P_G(\{\alpha \})$ and $Q_G(P,
\{\alpha \})$ on this particular $\alpha_i$. As we deal with polynomials of degree zero
or one in each $\alpha_{\ell}$ we write
$$P_G(\{\alpha \}) = a_i + b_i \alpha_i = b_i (a_i/b_i + \alpha_i ) \eqno(5.a)$$
$$\sum_C s_C \prod_{\ell \subset C} \alpha_{\ell} = d_i + e_i \alpha_i = e_i (d_i/e_i +
\alpha_i) \eqno(5.b)$$
\noindent where $a_i$, $b_i$, $d_i$ and $e_i$ are polynomials not depending explicitly
on $\alpha_i$. First, one can prove$^9$ that $a_i/b_i$ and $d_i/e_i$ are equal in the
limit where the number of propagators $I$ is infinite. Then, $Q_G(P, \{\alpha \}) =
e_i/b_i$ do not depend explicitly on $\alpha_i$ in this limit. However, due to the
constraint $h_0 = \sum\limits_i \alpha_i$, we have $\sum\limits_{j\not=i} \alpha_j =
h_0 - \alpha_i$ and $\alpha_i$ reappears each time $h_0$ does. And it does so each time
a power of $\bar{\alpha}_j$ appears because$^8$
$$\bar{\alpha}_{\ell} = 0(h_0/I) \eqno(6)$$
\noindent as suggested by remarking that $h_0^{I-1}/(I - 1) ! \sim (eh_0/I)^I$ leaving
a phase space $\sim eh_0/I$ for each $\alpha_{\ell}$. Gathering everything together we
get $F_G$ to depend on $\alpha_i$ through the integral$^8$
$$F_G = h_0(4 \pi)^{-dL/2} (\bar{b}_i)^{-d/2} \left [ Q_G (P, \{\bar{\alpha}\}) +
m^2h_0 \right ]^{-(I -d/2L)} \Gamma(I - d/2L) \cdot$$
$$\left [h_0^{I-2}/(I-2)! \right ] \int_0^{h_0} d\alpha (\mu_i + \alpha_i)^{-d/2}
\exp \left \{ - (I - d/2L) (1 - \beta ) \alpha_i/h_0 \right \} H_i(\alpha_i -
\bar{\alpha}_i) \eqno(7)$$
\noindent where we used the notation $\mu_i = \bar{a}_i/\bar{b}_i$ and $1 - \beta = [1
+ Q_G(P, \{\bar{\alpha}\})/(h_0m^2)]^{-1}$, $\bar{a}_i$, $\bar{b}_i$ and
$\{\bar{\alpha}\}$ being respectively $a_i$, $b_i$ and $\{\alpha \}$ where $\alpha_j$
is replaced by $\bar{\alpha}_j$ (for all $j \not= i$). $H_i(\alpha_i - \bar{\alpha}_i)$
represents the variation due to the dependence of the mean values $\bar{\alpha}_j$ on
$\alpha_i$ (not included in the replacement $h_0 \to h_0 - \alpha_i)$ when $\alpha_i$
is different from $\bar{\alpha}_i$ in $(\bar{b}_i)^{-d/2} [Q_G(P, \{\bar{\alpha}\}) +
m^2h_0]^{-(I-d/2L)}$. This factor which was not included in our first paper$^8$ does
not invalidate its conclusions concerning the validity of the mean-value approach and
the validity of (6). Its computation is, however, rather involved and will be
published elsewhere$^{10}$. Because of the exponential in (7), we can write
$$\bar{\alpha}_i = \bar{g}(\mu_i, H_i) (h_0/L + 1)) [1 + Q_G(P, \{
\bar{\alpha}\})/(h_0m^2)] \eqno(8)$$
\noindent where $I - d/2L$ has been replaced by $L + 1$ for $d = 4$ and $\phi^3$. Then,
$\bar{g}(\mu_i, H_i)$ is a constant to be determined by the equation (7). All what we
have said until now are general considerations. We have to specialize those for ladder
graphs. In the limit where $L \to \infty$ we get$^{11}$
$$P_G(\{\bar{\alpha}\}) = (\bar{\alpha}_-)^L \exp (yL) \ f(y) \eqno(9.a)$$
$$f(y) = {1 \over 2} y (1 + y^{-1})^2 \eqno(9.b)$$
\noindent with $y = (2\bar{\alpha}_+/\bar{\alpha}_-)^{1/2}$, $\bar{\alpha}_+$ being the
mean-value of the $\bar{\alpha}_{\ell}$'s for the propagators parallel to the ladder
and $\bar{\alpha}_-$ being the mean-value of the $\bar{\alpha}_{\ell}$'s for the
central propagators joining the two sides of the ladder. In the same limit we
get$^{11}$
$$Q_G(P, \{\bar{\alpha}\}) = (t/2) \ L \ \bar{\alpha}_+ + s \ \bar{\alpha}_- \exp (- yL)
[f(y)]^{-1} \ \ \ . \eqno(10)$$
\noindent So, putting the value for $P_G(\{\bar{\alpha}\})$ in (9.a) with
$\bar{\alpha}_-$ given by (8) with $\bar{g}(\mu_{\ell}, H_i) = \bar{g}_-$ we get (putting
back also the factor for the coupling constant $(- \gamma)^{2L+2}$) using asymptotic
expressions for factorials
$$F_G = (e^2/\sqrt{3}) [- \gamma /(m f(y))]^2 [- \gamma e/(m4 \pi 3 \sqrt{3})]^{2L}
\cdot $$ $$[\exp (- yL)/\bar{g}_- ]^{2L} [1 + Q_G(P,
\{\bar{\alpha}\})/(h_0m^2)]^{-(3L+1)} \ \ \ . \eqno(11)$$
Now let us sum over $L$ the ladder amplitudes in (11) and find a saddle-point. Then,
the saddle point equation reads
$$2 \ \ell n \ C^{st} + 3 \ \ell n (1 - \beta ) + (3L + 1) \left [ y + {1 \over L+1}
\right ] bs/(1 - \beta) = 0 \eqno(12)$$
\noindent with :
$$C^{st} = [- \gamma e/(m4 \pi 3 \sqrt{3})] [\exp (- y/\bar{g}_-] \eqno(13.a)$$
$$\beta = a+ b s \eqno(13.b)$$
$$a = (t \bar{g}_-/(2m^2)) (\bar{\alpha}_+ / \bar{\alpha}_-) \eqno(13.c)$$
$$bs = [\bar{g}_-/(m^2 f(y))] \ s \ \exp (-yL)/(L + 1) \ \ \ . \eqno(13.d)$$
\noindent We readily note that as $s \to \infty$, no solution exists for finite $L$.
This is consistent with the $L \to \infty$ limit we have used until now. \par

In the same way $bs$ tending to a constant yields no solution. If $bs \to \infty$ we get
$F_G$ to explode like $(- bs)^I$. The only possibility left is $bs$ tending to zero. In
that case $\beta$ tends to $a$ and we get
$$(1 - a) [(2/3) \ell n \ C^{st} + \ell n (1 - a)] + L \ y \ bs = 0 \eqno(14)$$
\noindent yielding the condition
$$s \exp (- y L_{sp}) = s_0 \quad , \qquad s_0 \ \hbox{constant} \eqno(15.a)$$
\noindent with
$$- (1 - a) [(2/3) \ell n \ C^{st} + \ell n (1 - a) ] = s_0 \ y \ \bar{g}_-/(m^2 f(y)) \
\ \ .  \eqno(15.b)$$
Thus, the saddle-point value for $L$, $L_{sp} = y^{-1} \ell n (s/s_0)$ which is
{\it a very natural condition} because it means that the dominant ladders have a
length which is proportional to the ``rapidity'' $\ell n (s/s_0)$ which has a very
important role in high energy physics. Replacing $L$ by $L_{sp}$ in (11) gives the
leading Regge trajectory immediately
$$\alpha (t/m^2) = y^{-1} [2 \ \ell n \ C^{st} + 3 \ \ell n (1 - \beta )] \ \ \ .
\eqno(16)$$ \noindent The trajectory $\alpha (0)$ can be computed if we know
$\bar{\alpha}_+$ and $\bar{\alpha}_-$. In fact, we do not compute $\bar{\alpha}_+$ and
$\bar{\alpha}_-$ themselves which are $0(1/L)$, (see (8)), but the finite quantities
$$\bar{\alpha}_- \Lambda = \bar{g}_-(\mu_-, H_i) \eqno(17.a)$$
$$\bar{\alpha}_+ \Lambda = \bar{g}_+ (\mu_+ , H_i) \eqno(17.b)$$
\noindent with the scale $\Lambda$ defined by
$$\Lambda = [(L + 1)/h_0] (1 - \beta ) \ \ \ . \eqno(18)$$
\noindent Then, $y = (2\bar{\alpha}_+ \Lambda/(\bar{\alpha}_- \Lambda ))^{1/2}$ is
also known. By inspection (13) also shows that $\beta$ is known once
$\bar{\alpha}_-\Lambda$ and $\bar{\alpha}_+\Lambda$ are. The fact that $F_G$ is
expressed in two ways in (4) and (7) allows to have an equation for each
$\bar{\alpha}_i$. Here, we have only two different $\bar{\alpha}_i$ 's,
$\bar{\alpha}_+$ and $\bar{\alpha}_+$ which give a total of two coupled equations. For
a given $\bar{\alpha}_i$ we have the equation
$$\exp (-(2 + \beta )/3) [I/(h_0\Lambda )] \int_0^{\infty} d(\alpha_i \Lambda ) [(\mu_i
\Lambda + \bar{\alpha}_i \Lambda )/(\mu_i \Lambda + \alpha_i \Lambda )]^{-d/2} \cdot$$
$$\exp
(- \alpha_i \Lambda ) H_i(\bar{\alpha}_i - \bar{\alpha}_i) = 1 \eqno(19)$$
\noindent In the particular case of ladders $\mu_+$ and $\mu_-$ are known functions of
$\bar{\alpha}_+$ and $\bar{\alpha}_-$ through$^{11}$ $1 + \bar{\alpha}_-/\mu_- =
y^{-1}$ and $1 + \mu_+/\bar{\alpha}_+ = 2y^{-1}$. The factor $\exp (-(2 + \beta )/3)$
takes into account the fact that in (4) $\bar{\alpha}_i = 0(h_0/I)$ whereas in (7)
$\bar{\alpha}_i = 0(h_0/(I - 1))$ because the mean-values are taken with only $(I - 1)$
variables instead of $I$ in (4). A more careful examination of the equations (19) (and in particular how to
determine $H_i(\bar{\alpha}_i - \bar{\alpha}_c))$ will be
published elsewhere$^{10}$. \par

We can now look at the solution for $\alpha (t)$ given in terms of $\ell n \ \gamma_m$
where $\gamma_m = \gamma e/(m 4 \pi 3 \sqrt{3})$ (in Minkowski space where $\gamma >
0$). \par 

In fig. 1 we displayed the Regge trajectory $\alpha (t/m^2)$ (in Minkowski space, $t$
being multiplied by $- 1$ relative to the Euclidean space) for two values of $\ell n
\ \gamma_m$, 0 (upper curves) and - 0.1 (lower curves). Due to the imperfect knowledge of
$H_i(\alpha_i - \bar{\alpha}_i)$ which can be represented in at least two different ways
we (at present) have an uncertainty which is shown by the magnitude of the difference
between the continuous curve and the dotted curve (for a given $\gamma_m$). The squares
show the values of the intercept $\alpha (0)$ as can be calculated from a
Bethe-Salpeter calculation$^{4,12}$ assuming $m = 0$ for the central rungs of the
ladders. We show the results obtained for $- 3.6 \leq t/m^2 \leq 0.8$. For $t/m^2$
lower than - 4 we have problems with our algorithm solving (19). For $t/m^2$ larger
than 0.8 we have a big uncertainty which is correlated with, we think, the solution
becoming complex well below the expected threshold at $t = 4m^2$. Maybe a new type of
anomalous threshold (of saddle point type~?$^{13}$) is showing up or, even simpler, our
method plainly fails to give meaningful results for large positive $t/m^2$. We note
that the shapes obtained are similar to those calculated$^{14}$ for the Pomeron using
the Lipatov equation. In fig. 2 the intercept $\alpha (0)$ is displayed as function of
$\gamma_m$ for $- 0.5 < \ell n \ \gamma_m < 0.4$. Again the uncertainty is shown by two
curves, plain and dotted. \par

We have a good agreement for $\alpha (0) \gsim 0.3$ with the Bethe-Salpeter
calculation$^{4,12}$. For lower values of $\alpha (0) \ \gamma_m$ becomes small and the
contribution outside the saddle-point starts to dominate. In particular we know that for
$\gamma_m \to 0$, the diagrams with a finite number of rungs will dominate the
contribution to the amplitude, giving an intercept tending to -~1. For large values of
$\ell n \ \gamma_m$, again, we encounter larger errors and so we didn't display the
results there. We hope to extend these calculations in the framework of QCD.

\vfill \supereject \centerline{\bf References} \bigskip
 \item{1 -} J. C. Polkinghorne, J. Math. Phys. {\bf 4}, 503 (1962). 
 
\item{2 -} B. W. Lee and R. F. Sawyer, Phys. Rev. {\bf 127}, 2266 (1962). 

\item{3 -} L. Bertocchi, S. Fubini and M. Tonin, Nuovo Cimento {\bf 25}, 626 (1962)~;
D. Amati, A. Stanghellini and S. Fubini, Nuovo Cimento {\bf 26}, 6 (1962). 

\item{4 -} N. Nakanishi, Phys. Rev. {\bf B135}, 1430 (1964). 

\item{5 -} N. I. Ussyukina and A.I. Davydychev, Phys. Lett. {\bf B305}, 136 (1993). 

\item{6 -} Ya. Ya. Balitskii, L.N. Lipatov, Sov. J. Nucl. Phys. {\bf 28} (1978) 822. 
\item{} R. Kischner, L.N. Lipatov, Z. Phys. {\bf C45} (1990).
\item{} R.E. Hancock, D.A. Ross, Nucl. Phys. {\bf B383}, 575 (1992).

\item{7 -} R. Gastmanns, W. Troost and T.T. Wu, Nucl. Phys. {\bf 365}, 404 (1991). 

\item{8 -} R. Hong Tuan, Phys. Lett. {\bf 286}, 315 (1992). 

\item{9 -}R. Hong Tuan, Factorization of Spanning Trees on Feynman Graphs, preprint
LPTHE Orsay 92/59 (dec. 1992). 

\item{10 -} R. Hong Tuan, to be published. 

\item{11 -} R. Hong Tuan, to be published. 

\item{12 -} Multiperipheral Dynamics, M.L. Goldberger in Subnuclear Phenomena, 1969
International School of Physics ``Ettore Majorana''. Editor A. Zichichi, Academic Press
(1970).

\item{13 -} Quantum Field Theory, C. Itzykson and J. B. Zuber, p. 307, Mc Graw-Hill
(1980).

\item{14 -} R.E. Hancok and D.A. Ross, Nucl. Phys. {\bf B394} (1993) 200.

\vfill\supereject
\centerline{\bf Figure Captions} \bigskip
{\parindent=1 cm
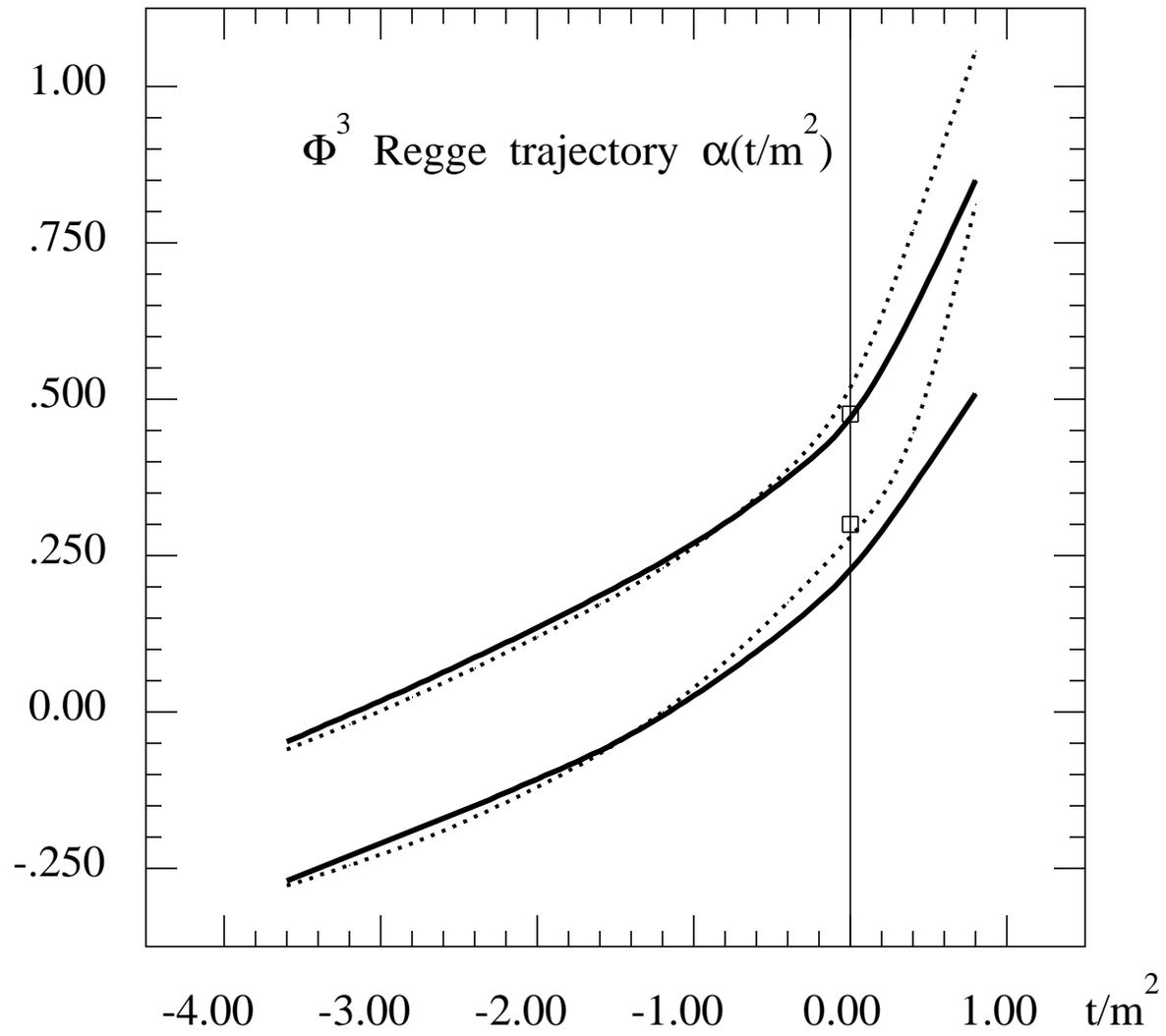
\item{\bf Fig. 1} The Regge trajectory $\alpha (t/m^2)$ as displayed for two values of
$\ell n \ \gamma_m$, 0. for the upper curves and - 0.1 for the lower curves. The size
of the uncertainty is given by the difference between the corresponding plain and
dotted curves. Squares indicate the intercept $\alpha (0)$ obtained from the
Bethe-Salpeter equation$^{4,12}$. 
  
\vskip 3 mm

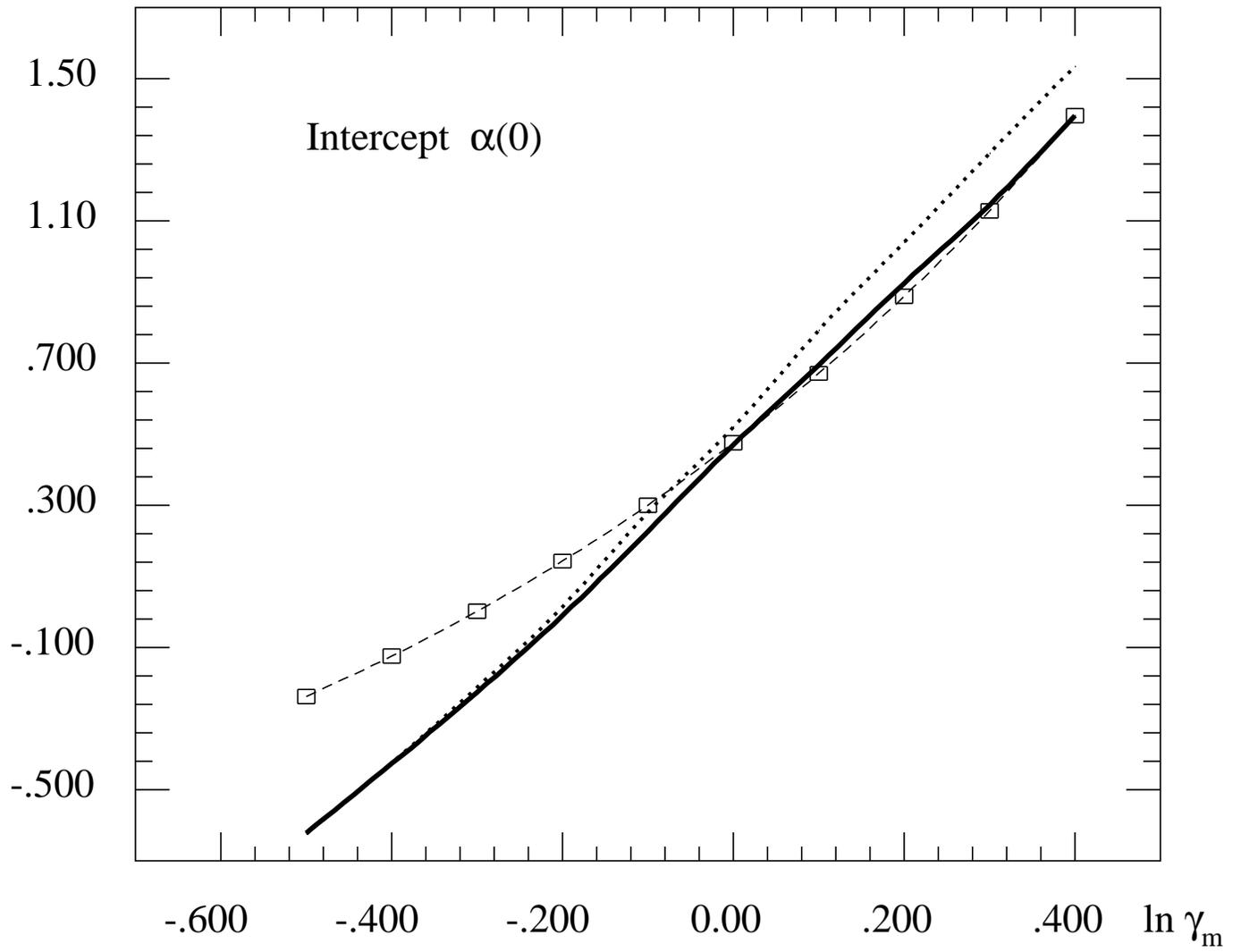
\item{\bf Fig. 2} The intercept $\alpha (0)$ is shown for the range $-.5 \leq \ell n \
\gamma_m \leq .4$. Again, the uncertainty is given by the difference between the plain
and the dotted curves. The Bethe-Salpeter result is also displayed with the dashed
line supporting empty squares.  \vskip 3 mm

\par}

\bye